\documentclass[onecolumn,preprintnumbers,pra,superscriptaddress]{revtex4}

\usepackage{amssymb,amsmath,graphicx,dcolumn,bm,float} 
\usepackage[utf8]{inputenc}
\usepackage[T1]{fontenc}
\usepackage{xcolor}
\usepackage[colorlinks,linkcolor=black,citecolor=blue]{hyperref}

\begin{document}

\title{$\mathcal{PT}-$symmetry and chaos control via dissipative optomechanical coupling}
\author{S. R. Mbokop Tchounda}
\email{rolande.mbokop@facsciences-uy1.cm}
\affiliation{Department of Physics, Faculty of Science, University of Yaounde I, P.O. Box 812, Yaounde, Cameroon}

\author{P. Djorwé}
\email{djorwepp@gmail.com}
\affiliation{Department of Physics, Faculty of Science, 
University of Ngaoundere, P.O. Box 454, Ngaoundere, Cameroon}
\affiliation{Stellenbosch Institute for Advanced Study (STIAS), Wallenberg Research Centre at Stellenbosch University, Stellenbosch 7600, South Africa}

\author{M. V. Tchakui}
\email{muriellevan@yahoo.fr}
\affiliation{Department of Electrical and Electronic Engineering, National Higher Polytechnic Institute, University of Bamenda, PO Box 39, Bamenda, Cameroon}

\author{S. G. Nana Engo}
\email{serge.nana-engo@facsciences-uy1.cm}
\affiliation{Department of Physics, Faculty of Science, University of Yaounde I, P.O. Box 812, Yaounde, Cameroon}

\begin{abstract} 
We study a dissipative mechanically coupled optomechanical system that hosts gain and loss. The gain (loss) is engineered by driving a purely dispersive optomechanical cavity with a blue-detuned (red-detuned) electromagnetic field. By taking into account the dissipative coupling, the Exceptional Point (EP), which is the $\mathcal{PT}-$symmetry phase transition, occurs at low threshold driving strength compared to what happens in a solely dispersive system. In the linear regime, the $\mathcal{PT}-$symmetry is unbroken and the dissipative term induces strong coupling between the mechanical resonators, leading to an increase of energy exchange. For strong enough driving, the system enters into a nonlinear regime where the $\mathcal{PT}-$symmetry is broken. In this regime, the mechanical resonators exhibit chaotic beats like-behaviour in the purely dispersive system. By switching on the dissipative coupling, the complex dynamics is switched off, and this restores regular dynamics into the system. This work suggests a way to probe quantum phenomena in dissipative $\mathcal{PT}-$symmetric systems at low-threshold driving strength. It also provides a new control scheme of complex dynamics in optomechanics and related fields.
\end{abstract}

\maketitle

\date{\today}


%
\section{Introduction} \label{Intro}

Optomechanical coupling, an interaction between an electromagnetic field and mechanical objects, is a nice platform to foster quantum breakthroughs for both fundamental and technological purposes \cite{Aspelmeyer_2014}. Two main optomechanical coupling are investigated in the literature, the dispersive couplings \cite{Aspelmeyer_2014} and the dissipative/reactive coupling \cite{Li_2009,Huang_2010,Fu_2012,Weiss_2013,Gu_2013,Qu_2015,Wu_2014,Tagantsev_2018}. The former is well known and widely used, while the latter, which also leads to nice features as well seems to attract less attention. The process where the cavity frequency is modulated by the mechanical vibrations leads to the dispersive coupling, while the dissipative coupling results from the modulation of both the resonance frequency and the linewidth of the cavity mode by the mechanical vibrations. In the dispersive case, optomechanical cavities (COM) have led to several investigations including quantum ground state cooling \cite{Clark_2017,Djorw__2012}, quantum entanglement \cite{Kotler_2021,Tchodimou_2017,Djorw__2014}, squeezing \cite{Wollman_2015,Djorwe2013} , nonlinear dynamics \cite{Djorwe2013,Alphonse_2022,Djorwe2019,Djorwe_2020} and chaos \cite{Monifi_2016,Djorwe_2018,Kingni_2020}. In terms of array systems, dispersive COM have been used to engineer Parity-Time ($\mathcal{PT}$) symmetry and exceptional points (EPs), which are Non-Hermitian degeneracies. These EPs have triggered interesting and intriguing effects in physics EPs \cite{Miri_2019}. Among them are stopping light \cite{Goldzak_2018}, loss-induced suppression and revival of lasing, pump-induced lasing death, unidirectional invisibility \cite{Peng_2014}, collective phenomena \cite{Djorwe_2020,Djorwe_2018} , sensors \cite{Djorwe_2019,Chen_2017,Hodaei_2017} and topological features \cite{Xu_2016,Ren_2022,Peano_2015}. Similarly, interesting achievements have been made in the dissipative's COM framework. These achievements include normal mode splitting \cite{Huang_2010}, strong-coupling effect \cite{Weiss_2013}, squeezing \cite{Gu_2013,Qu_2015,Tagantsev_2018} and detection \cite{Wu_2014}.  To the best of our knowledge, there are no $\mathcal{PT}-$symmetry nor EPs studies  in the literature involving dissipative COM, although they share common promising features with their dispersive counterparts. Addressing and bridging this gap may provide new opportunities for optomechanical platforms to uncover interesting features.  

Our work aims to investigate the effect of dissipative coupling on $\mathcal{PT}-$ symmetry and EPs engineering in coupled optomechanical cavities. The benchmark system consists of two optomechanical systems where the involved mechanical resonators are mechanically coupled. The gain (loss) in the system is controlled by driving the resonator with a blue (red) detuned laser. More interestingly, the dissipative coupling rate in our proposal has been revealed as an additional parameter to adjust the gain (loss) rate even at a fixed driving field. This flexibility to control gain and loss is a good prerequisite for $\mathcal{PT}-$symmetry engineering, which is an advantage over the dispersive COM structures. In the linearized regime, we carried out the linear stability study and confirmed that this stability fails at the EP when the system reaches the phonon lasing or the parametric gain threshold \cite{Foulla_2017}. For purely dissipative coupling, we found that the EP is shifted towards low values of the driving field as the dissipation rate increases. This leads to strong coupling and energy exchange between the hybridized supermodes in the system, even for weak driving strength. In the nonlinear regime, the dissipative coupling quenches the complex dynamics by switching them to regular behavior, providing a chaos control scheme. Due to these interesting features, the strong coupling effect induced here can be deeply explored for quantum investigations, while the chaos control may offer great opportunities for collective phenomena \cite{Djorwe_2020} and dynamical entanglement \cite{Wang_2014}.  

This work is organized as follows. The \autoref{sec:MoEq} describes the model and its dynamical equations up to the EP features. The \autoref{sec:li_dyn} is devoted to the dissipatively induced strong coupling while the chaos control through dissipative coupling is presented in \autoref{sec:ctr_chaos}. The work is concluded in \autoref{sec:concl}. 

\section{Modelling and dynamical equations} \label{sec:MoEq}

\subsection{Modelling}

\begin{figure}[tbh]
  \centering
  \includegraphics[width=0.9\linewidth]{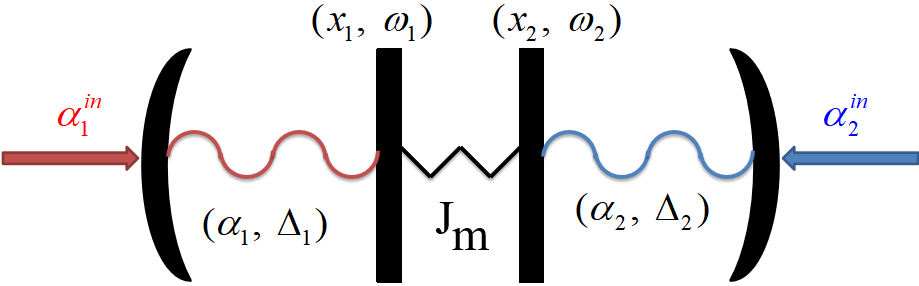}
  \caption{Optomechanical cavities containing both dispersive and dissipative couplings. The mechanical resonators are mechanically coupled.}
  \label{fig:Fig1}
  \end{figure}

Our benchmark system consists of two optomechanical cavities in which the mechanical resonators are mechanically coupled as sketched in FIG.~\ref{fig:Fig1}. One of the COM is driven by the red-detuned laser to engineer losses in the cavity, while the other is driven by a blue-detuned field to provide gain to the cavity with that offsets the losses. This system is assumed to contain both dispersive and dissipative couplings.  

In the rotating frame of the driving fields ($\omega_p^j$ with $j=1,2$), the Hamiltonian ($\hslash=1$) describing this system is  
\begin{equation}\label{eq:Hamil1}
H=H_{\rm OM}+H_{\rm int}+H_{\rm drive}+H_{\kappa}+H_{\gamma_m},
\end{equation}  
where 
\begin{align}\label{eq:Hamil2}
H_{\rm OM}&=\sum_{j=1,2}\left[-\Delta_j a^{\dag}_j a_j+\omega_j b^{\dag }_jb_j +\frac{L}{c}\bar{n}_g\omega_p^j(a^{in}_j)^2\right],\\
H_{\rm int}&=-\sum_{j=1,2}\left[g_ja^{\dag}_ja_j(b^{\dag}_j+b_j)\right]-J_m (b^{\dag }_{1}b_{2}+b_{1}b^{\dag }_{2}),\\
H_{drive}&=i\sqrt{\kappa_x}\alpha^{\rm in}(a^{\dag}_j-a_j).  
\end{align}
where $H_{\rm OM}$, $H_{\rm int}$ and $H_{\rm drive}$ are the optomechanical, coupling interactions and the driving Hamiltonians respectively. The cavity and the mechanical decay are captured by the Hamiltonians $H_{\kappa}$ and $H_{\gamma_m}$. The annihilation bosonic field operators describing the optical and mechanical resonators are $a_j$ and $b_j$. The mechanical displacements $x_j$ are related to the operators $b_j$ as $x_j=x_{\rm ZPF}(b_j+b^{\dag}_j)$, where $x_{\rm ZPF}=\sqrt{\frac{\hslash}{2m\omega_j}}$ is the zero-point fluctuation amplitude of the mechanical resonator, where $\omega_j$ being the mechanical frequency of the $j^{th}$ resonator as shown in FIG.~\ref{fig:Fig1}. The frequency detunings between the driving ($\omega_p^j$) and the cavity ($\omega_{cav}^j$) are defined by $\Delta_j=\omega_p^j-\omega_{cav}^j$, and the phonon-hopping rate between the two mechanical resonators is $J_m$. $L$ is the length of the free-standing waveguide interacting with the optical cavity, $c$ the speed of light in vacuum, and $\bar{n}_g$ the group 
index of the waveguide optical mode. The amplitude of the driving fields is $\alpha^{in}$, which is assumed to be the same for both cavities, and is related to the input power $P_{in}$ as  $a^{in}=\sqrt{\frac{P_{in}}{\hbar\omega_p}}$.  The dispersive optomechanical rates are $g_j$ while the dissipative coupling $g_{\kappa}$ results from the modulation of the cavity dissipation $\kappa_{x}$ by the mechanical displacement as, 
\begin{equation}\label{eq:eq2} 
\kappa_{x}\approx\kappa+\kappa_{\rm om}x=\kappa+g_{\kappa}(b^{\dag}+b)=\kappa(1+\eta(b^{\dag}+b)), 
\end{equation}
where $g_{\kappa}=\kappa_{\rm om}x_{\rm ZPF}$ and 
$\eta=\frac{g_{\kappa}}{\kappa}$.

\subsection{Dynamical equations}

From the Heisenberg's equation $\dot{\mathcal{O}}=i[H,\mathcal{O}]+\mathcal{N}$, where $\mathcal{O}\equiv (a_j,b_j)$ and $\mathcal{N}\equiv (a^{in}_j,b^{in}_j)$ the related noise operator, the quantum Langevin equations (QLEs)  of our system read,
\begin{subequations}\label{eq:Nldyna}
\begin{align}
\begin{split}
\dot{a_j}&=\left(i\left(\Delta_j+(g_j+i\frac{\eta_j \kappa}{2}) (b^{\dag }_j+b_j)\right)-\frac{\kappa}{2}\right)a_j \\  &+\sqrt{\kappa}\left(1+\frac{\eta_j}{2}(b^{\dag}_j+b_j)\right)\alpha^{in}+\sqrt{\kappa}a^{in},
\end{split}\\ 
\begin{split}
\dot{b_j}&=-(i\omega_j +\frac{\gamma_m}{2})b_j +i J_m b_{3-j} \\ 
&+\eta_j\sqrt{\kappa}(a^{\dag }_j-a_j)\alpha^{in}_j+ig_j a^{\dag}a + \sqrt{\gamma_m}b^{in}_j. 
\end{split}
\end{align}
\end{subequations}
These noise operators $\mathcal{N}\equiv (a^{in}_j,b^{in}_j)$ have zero mean and are characterized by the following autocorrelation  functions \cite{Tchodimou_2017},
\begin{align}\label{eq:noise}
&\langle \mathcal{N}(t) \mathcal{N} ^{\dag} (t^{\prime}) \rangle &=(n_{\nu}+1)\delta(t-t^{\prime}),\\
&\langle \mathcal{N}^{\dag}(t) \mathcal{N}(t^{\prime}) \rangle &=n_{\nu}\delta(t-t^{\prime}),
\end{align}
with $n_{\nu}\equiv (n_{th},n_a)$, where  $n_{th}=\rm {exp(\tfrac{\hslash \omega_j}{k_B T}-1)^{-1}}$ and $n_a=\langle a^{in \dag}_j a^{in}_j \rangle$. For a seek of simplicity and without loss of generality, we assume in the following that $\omega_j\equiv \omega_m$, $g_j\equiv g_m$ and $\eta_j \equiv \eta$.

To gain insight into our system and to understand its dynamics, we linearize the set of nonlinear equations given in Eq.~\eqref{eq:Nldyna}, which consists of splitting the field operators as $\mathcal{O}=\langle O \rangle +\delta \mathcal{O}$, where $\langle O \rangle\equiv\beta_j,\alpha_j$ are the coherent complex parts of the operator and $\delta \mathcal{O}\equiv\delta\beta_j,\delta\alpha_j$ their associated fluctuations. Therefore, the coherent dynamics reads
\begin{align}\label{eq:cldyn}
\begin{split}
&\dot{\alpha_j}=(i\tilde{\Delta_j}-\frac{\kappa}{2})\alpha_j +\sqrt{\kappa}(1+\eta\Re{(\beta_j)})\alpha^{in},\\
&\begin{aligned}
\dot{\beta_j}=&-(i\omega_m+\frac{\gamma_m}{2})\beta_j +iJ_m\beta_{3-j}\\
&+\eta\sqrt{\kappa}(\alpha^{\dag}_j-\alpha_j)\alpha^{in}+ig \left |\alpha_j\right|^2, 
\end{aligned}
\end{split}
\end{align}
and the fluctuation dynamics yields,
\begin{equation}\label{eq:qdyn}
\begin{split}
\dot{\delta\alpha_j}&=(i\tilde{\Delta_j}-\frac{\kappa}{2})\delta\alpha_j +\sqrt{\kappa}\left(1+\eta\Re{(\beta_j)}\right)\delta\alpha^{in} \\&+ \left(i\alpha_j(g+i\frac{\eta\kappa}{2})+\frac{\eta\sqrt{\eta}\alpha^{in}}{2}\right)(\delta\beta_j+\delta\beta^{in}_j),\\
\dot{\delta\beta_j}&=-(i\omega_m+\frac{\gamma_m}{2})\delta\beta_j +iJ_m\delta\beta_{3-j}\\
&+\eta\sqrt{\kappa}(\delta\alpha^{\dag}_j-\delta\alpha_j)\alpha^{in}+\eta\sqrt{\kappa}(\alpha^{\dag}_j-\alpha_j)\delta\alpha^{in}\\
&+ ig(\alpha^{\dag}_j\delta\alpha_j+\alpha_j\delta\alpha^{\dag}_j), 
\end{split} 
\end{equation}
with $\tilde{\Delta_j}=\Delta_j+2\Re{(\beta_j)}(g+i\frac{\eta\kappa}{2})$.

In the following we will work out the steady state equation of our system from Eq.~\eqref{eq:cldyn} and use it to study both the stability and to derive the EP. The steady state condition is that the averaged variables in Eq.~\eqref{eq:cldyn} are no longer time dependent. This means that $\dot{\alpha_j}=\dot{\beta_j}=0$, which leads to,
\begin{equation}\label{eq:c0dyn}
\begin{split}
&0=(i\tilde{\Delta_j}-\frac{\kappa}{2})\bar{\alpha_j} +\sqrt{\kappa}(1+\eta\Re{(\bar{\beta_j)}})\alpha^{in},\\
&\begin{split}
0&=(i\omega_m+\frac{\gamma_m}{2})\bar{\beta_j} -iJ_m\bar{\beta}_{3-j}\\
&-\eta\sqrt{\kappa}(\bar{\alpha}^{\dag}_j
-\bar{\alpha}_j)\alpha^{in}-ig \left |\bar{\alpha}_j\right|^2.
\end{split}
\end{split}    
\end{equation}

This set of equations can be solved to obtain the steady state equation for either $\bar{\alpha_j}$ or  $\bar{\beta_j}$. After tedious computation, and because the expressions are cumbersome, we  present only the equation for $\bar{\alpha_j}$ which reads,
\begin{equation}\label{eq:stst}
\left|\bar{\alpha_j}\right|^6+a_0\left|\bar{\alpha_j}\right|^5+a_1\left|\bar{\alpha_j}\right|^4+a_2\left|\bar{\alpha_j}\right|^3+a_3\left|\bar{\alpha_j}\right|^2+a_4\left|\bar{\alpha_j}\right|+a_5=0,
\end{equation}
where the involved coefficients are, $a_0=\frac{b_0}{c}$, $a_1=\frac{b_1}{c}$, $a_2=\frac{b_2}{c}$, $a_3=\frac{b_3}{c}$, $a_4=\frac{b_4}{c}$ and $a_5=\frac{b_5}{c}$ with
\begin{align}
\begin{split}
&b_0=-4g\Omega^2\eta\sqrt{\kappa}\alpha^{in}(\eta^2\kappa^2+4),\\
\begin{split}
&b_1=4\Omega^2\eta^2\kappa\alpha^{in}(\eta^2\kappa^2+4)-\kappa(g\Omega\eta\alpha^{in})^2\\
&+g\Omega(4\Delta_j+\eta\kappa^2), 
\end{split}\\
&b_2=4\Omega^2g(\eta\alpha^{in})^3\kappa\sqrt{\kappa}-2\Omega\sqrt{\kappa}\alpha^{in}(4\Delta_j+\eta\kappa^2),\\
&b_3=2\Omega\eta\kappa g(\alpha^{in})^2-(2\Omega\kappa)^2(\eta \alpha^{in})^4+\Delta_j^2+\frac{\kappa^2}{4},\\
&b_4 =-4\eta\kappa (\alpha^{in})^3,\\
&b_5 =-\kappa \alpha^{in},
\end{split}    
\end{align}
and the other of the parameters are, $c=(g\Omega)^2((\eta\kappa)^2+4)$ and $\Omega=\tfrac{(\omega_m^2-J_m^2)(\omega_m+J_m)}{(J_m^2-\omega_m^2)^2+(\omega_m\gamma_m)}$. The solution of Eq.~\eqref{eq:stst} can then be used in the fluctuation dynamics Eq.~\eqref{eq:qdyn} for the EP analysis. For this purpose we change to another interaction picture by introducing the following slowly varying operators with tildes, $\delta a=\delta \tilde{a}e^{i\tilde{\Delta}t}$ and $\delta b_j=\delta \tilde{b_j}e^{-i\omega_j t}$.  In the limit of $\omega_j\gg(G_j,\kappa)$ the rotating wave approximation can be invoked depending on the blue or red sideband resonances, where the fast oscillating terms are neglected. Moreover, under the condition $\kappa\gg(G,\gamma_j)$, which is satisfied in our analysis, the cavity field can be eliminated adiabatically, and this leads us to the effective dynamical system for the two mechanical resonators \cite{Jiang_2021}:\begin{equation}\label{eq:effeq}
\begin{split}
\dot{\delta\beta_1}&=(\frac{\gamma_{eff}^1}{2}-i\omega_{eff}^1)\delta\beta_1+iJ_m \delta\beta_2
 +\sqrt{\gamma_m}\delta\beta^{in}_1\\ 
 &+(i\sqrt{\Gamma}+2\eta\alpha^{in})(1+\eta\Re{(\bar{\beta})})\delta\alpha^{in\dag},\\ 
\dot{\delta\beta_2}&=(\frac{\gamma_{eff}^2}{2}-i\omega_{eff}^2)\delta\beta_2+iJ_m \delta\beta_1
 +\sqrt{\gamma_m}\delta\beta^{in}_2\\ 
 &+(i\sqrt{\Gamma}-2\eta\alpha^{in})(1+\eta\Re{(\bar{\beta})})\delta\alpha^{in}.\\ 
\end{split}    
\end{equation}
$\Gamma=\tfrac{4|G|^2}{\kappa}$ is the optical damping with $G=g\bar{\alpha}$ the optomechanical induced coupling rate. We have assumed that the two cavities are driven with the same driving strength $\alpha^{in}_j\equiv\alpha^{in}$ so that $\bar{\alpha_j}\equiv\bar{\alpha}$ and $\bar{\beta_j}\equiv\bar{\beta}$. We have also defined both the effective dissipation and the frequency as,  
\begin{align}\label{eq:effp}
\begin{split}
\gamma_{eff}^1&=\Gamma-\gamma_m-2\eta^2\alpha^{in}(\alpha^{\ast}\sqrt{\kappa}-\alpha^{in}),\\
\gamma_{eff}^2&=-(\Gamma+\gamma_m)+2\eta^2\alpha^{in}(\alpha\sqrt{\kappa}-\alpha^{in}),\\
\omega_{eff}^1&=\omega_m+\frac{\eta\sqrt{\Gamma}}{2}(\alpha^{\ast}\sqrt{\kappa}+\alpha^{in}),\\
\omega_{eff}^2&=\omega_m+\frac{\eta\sqrt{\Gamma}}{2}(\alpha\sqrt{\kappa}+\alpha^{in}).
\end{split}
\end{align}

\subsection{EP analysis}

\begin{figure}[tbhp]
\begin{minipage}{\textwidth}
  \centering
  \resizebox{0.48\textwidth}{!}{
  \includegraphics{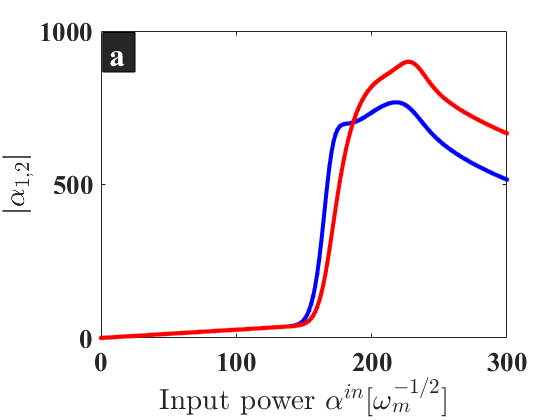}}
  \resizebox{0.48\textwidth}{!}{
  \includegraphics{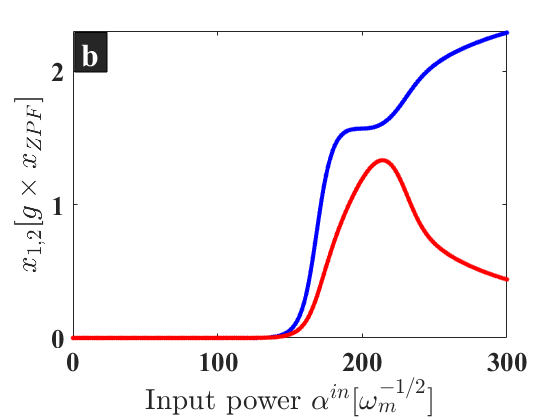}}
  \resizebox{0.48\textwidth}{!}{
  \includegraphics{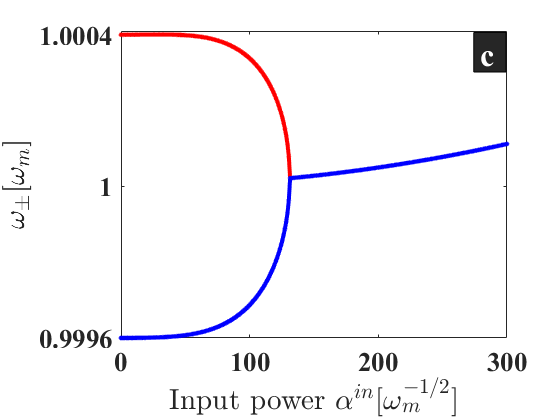}}
  \resizebox{0.48\textwidth}{!}{
  \includegraphics{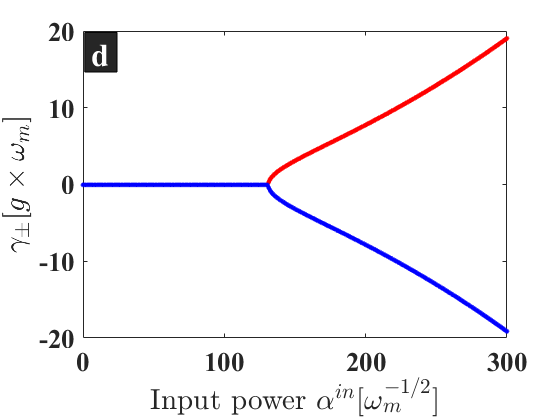}}
  \caption{(a) and (b) are the photon number and the mechanical displacement obtained from numerical simulation of the full nonlinear equation in Eq.~\eqref{eq:Nldyna}, where the noise terms are dropped and the operators are averaged, i.e., $\langle a_j \rangle \equiv \alpha_j$, and $\langle b_j \rangle \equiv \beta_j$. (c) and (d) are the eigenfrequencies and eigendampings of our system. It can be seen that they both coalesce at the EP, which agrees well with the parametric instability threshold (see (a) and (b)) where the system lost its stability. The parameters used are $\Delta_1=\omega_m$, $\Delta_2=-\omega_m$, $g_m=1.076\times10^{-4}\omega_m$, $\kappa=7.3\times 10^{-2}\omega_m$,  $\eta=0.1 g_m$, $\gamma_m=1.076\times 10^{-5}\omega_m$, and $J_m=4\times 10^{-4}\omega_m$.}
  \label{fig:Fig2}
  \end{minipage}
\end{figure}

In order to process with the EP analysis, drop the noise term in Eq.~\eqref{eq:effeq} and write the noiseless effective equations in the compact form,
\begin{equation}\label{eq:eff}
   i\partial_t \psi=\rm{H_{eff}}\psi,
\end{equation}
with $\psi=(\delta\beta_1, \delta\beta_2)$ and the effective Hamiltonian given by, 
\begin{equation}\label{eq:eff_H}
\rm{H_{eff}}=
\begin{pmatrix}
\omega_{eff}^1 + i\frac{\gamma_{eff}^1}{2} & -J_m \\
-J_m   & \omega_{eff}^2 + i\frac{\gamma_{eff}^2}{2}
\end{pmatrix}.
\end{equation}
The eigenvalues of the Hamiltonian given in Eq.~\eqref{eq:eff_H} are,
\begin{equation}\label{eq:Eig}
\lambda_{\pm}=\frac{1}{2}(\omega_{eff}^1 +\omega_{eff}^2)+\frac{i}{4}(\gamma_{eff}^1 +\gamma_{eff}^2) \pm \frac{\sigma}{4},
\end{equation}
where,  
\begin{equation}\label{eq:Sig}
\sigma=\sqrt{(2\Delta\omega_{eff}+i\Delta\gamma_{eff})^2+16J_m^2},
\end{equation}
with $\Delta\omega_{eff}=\omega_{eff}^1 - \omega_{eff}^2$ and $\Delta\gamma=\gamma_{eff}^1 -\gamma_{eff}^2$. 
From these eigenvalues, the eigenfrequencies and eigendampings of the system are defined as the real ($\omega_{\pm}=\Re{(\lambda_{\pm})}$) and imaginary ($\gamma_{\pm}=\Im{(\lambda_{\pm})}$) parts of $\lambda_{\pm}$ respectively. The EP occurs when these two pairs of frequencies and dampings coalesce, i.e. $\omega_{-}=\omega_{+}$ and $\gamma_{-}=\gamma_{+}$, which means that $\sigma=0$. This condition on $\sigma$ can be fulfilled in our proposal by tuning the driving field instead of $\eta$, which is not easily tunable as a driving field. Fig.~\ref{fig:Fig2} shows the agreement between the numerical stability and the analytical investigation by the EP. In fact, Fig.~\ref{fig:Fig2}(a-b) are the numerical simulations of the photon number and the mechanical displacement from Eq.~\eqref{eq:Nldyna}, where the noise terms are dropped and the operators are averaged, i.e., $\langle a_j \rangle \equiv \alpha_j$ and $\langle b_j \rangle \equiv \beta_j$. These bosonic quantities are extracted from the steady state time series for different values of the driving strength. The photon number is given by $\langle \alpha^{\dag}_j\alpha_j \rangle$ while the mechanical position is quantified by $x_j=\beta_j+\beta_j^{\dag}$. It can be seen that these quantities are exponentially amplified around $\alpha^{in}\sim 130\omega^{1/2}_{m}$, known as the parametric instability threshold where the system loses its stability \cite{Foulla_2017}. Furthermore, Figs.~\ref{fig:Fig2}(c-d) show the eigenfrequencies and eigendampings of our system, which both coalesce at the EP, and this also happens around the threshold mentioned in Figs.~\ref{fig:Fig2}(a-b). This agreement confirms the validity of our analytical investigation. From this analysis, we devote the next section to study the effect of the dissipative coupling $\eta$ on the dynamical behavior of our system.    

\section{Dissipation induces strong coupling} \label{sec:li_dyn}

\begin{figure}[tbh]
 \begin{minipage}{\textwidth}
  \begin{center}
  \resizebox{0.48\textwidth}{!}{
  \includegraphics{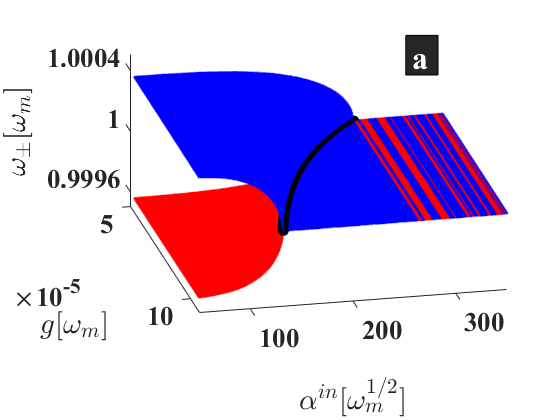}}
  \resizebox{0.48\textwidth}{!}{
  \includegraphics{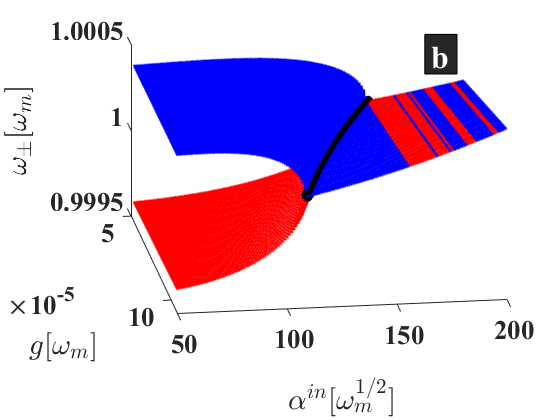}}
  \resizebox{0.48\textwidth}{!}{
  \includegraphics{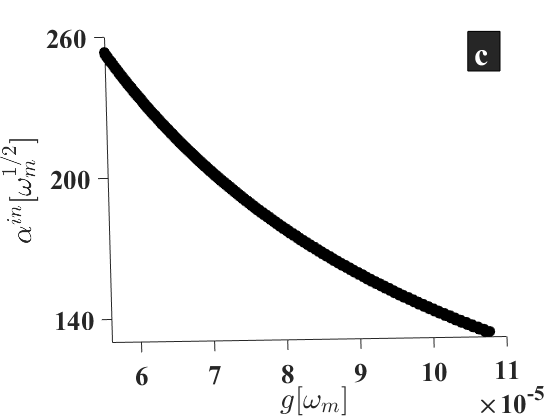}}
  \resizebox{0.48\textwidth}{!}{
  \includegraphics{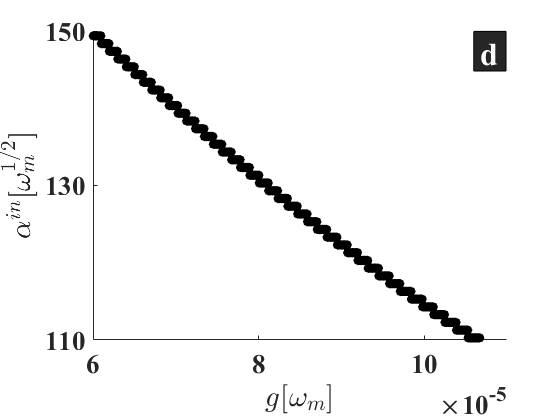}}
  \end{center}
  \caption{(a) and (b) are the 3D representation of the eigenfrequencies in purely dispersive ($\eta=0$) and with a fixed dissipative coupling rate at $\eta=g_m$. (c) and (d) highlight the EP evolution, which is depicted with black line in (a) and (b). The other parameters are the same as in Fig.~\ref{fig:Fig2}.}
  \label{fig:Fig3}
  \end{minipage}
\end{figure}

Due to the relationship between the effective parameters and the dissipative coupling in Eq.~\eqref{eq:effp}, it is obvious that the gain-loss engineering will depend on $\eta$, and this will induce a modification on the EP behavior in our system. Fig.~\ref{fig:Fig3}a shows the structure of the eigenfrequencies for the purely dispersive case ($\eta=0$), while Fig.~\ref{fig:Fig3}b shows the eigenfrequencies when the dissipative coupling is fixed at $\eta=g_m$. It can be seen that in the purely dispersive case the EPs appear for relatively high values of the driving strength compared to the case where the dissipative coupling is taken into account. This observation is highlighted in Fig.~\ref{fig:Fig3}c and Fig.~\ref{fig:Fig3}d, where the 2D projections of Fig.~\ref{fig:Fig3}a and Fig.~\ref{fig:Fig3}b are extracted.  Indeed, Fig.~\ref{fig:Fig3}d shows that for strong dispersive coupling, EP can be engineered for low values of the driving field when $\eta$ is involved.  It follows that dissipative coupling induces a low power $\mathcal{PT}-$symmetry phase transition compared to the purely dispersive coupling case. From an experimental point of view, this may be a crucial aspect to consider and a good motivation for more dissipative $\mathcal{PT}-$symmetry features. To confirm the above analysis, we have further performed numerical simulations in both the linear and nonlinear regime. From Fig.~\ref{fig:Fig2}, it is clear that the linear regime corresponds to the unbroken $\mathcal{PT}-$symmetry regime, where the system exhibits Rabi oscillation-like dynamics due to the two frequencies involved, as shown in Fig.~\ref{fig:Fig2}c. In the nonlinear regime, however, Fig.~\ref{fig:Fig2} shows that the system is in the broken $\mathcal{PT}-$symmetry phase, where the mechanical resonators dissipate differently, as shown in Fig.~\ref{fig:Fig2}d. 

Our numerical simulations for the linear regime are displayed in Fig.~\ref{fig:Fig4}. Starting from $\eta=0$ in Fig.~\ref{fig:Fig4}a for a fixed value of the driving strength $\alpha^{in}=20[\omega_m^{1/2}]$, one can see that the system is driven towards a strong coupling regime by tuning the value of $\eta$ (see Fig.~\ref{fig:Fig4}(b-d)). In fact, Fig.~\ref{fig:Fig4}a shows almost no Rabi oscillations, while these oscillations are clearly visible in Fig.~\ref{fig:Fig4}d, clearly showing the different frequencies in the system. This strong coupling is reminiscent of a strong energy exchange between the two mechanical resonators involved. It is concluded that the dissipative coupling rate $\eta$ can be used to enhance PT-symmetry features such as strong coupling \cite{Weiss_2013,Huang_2010} and to stimulate efficient energy transfer \cite{Xu_2016}.

\section{Dissipative control of chaos} \label{sec:ctr_chaos}

This section describes the dynamical behavior of our system above the nonlinear regime, i.e., above the parametric threshold shown in Fig.~\ref{fig:Fig2}. Above this threshold and in the purely dispersive regime, the mechanical resonators oscillate with a common frequency while each of them dissipates differently according to the eigenvalues depicted in Fig.~\ref{fig:Fig2}(c-d). In fact, these mechanical resonators will first settle into a limit cycle regime (or in regular oscillations) and they may exhibit collective phenomena \cite{Djorwe.2013,Djorwe_2020}. Once that the driving force is strong enough, multi-frequency and chaotic behavior can be induced \cite{Djorwe_2018}. 

\begin{figure}[tbh]
\centering
\includegraphics[width=1. \textwidth]{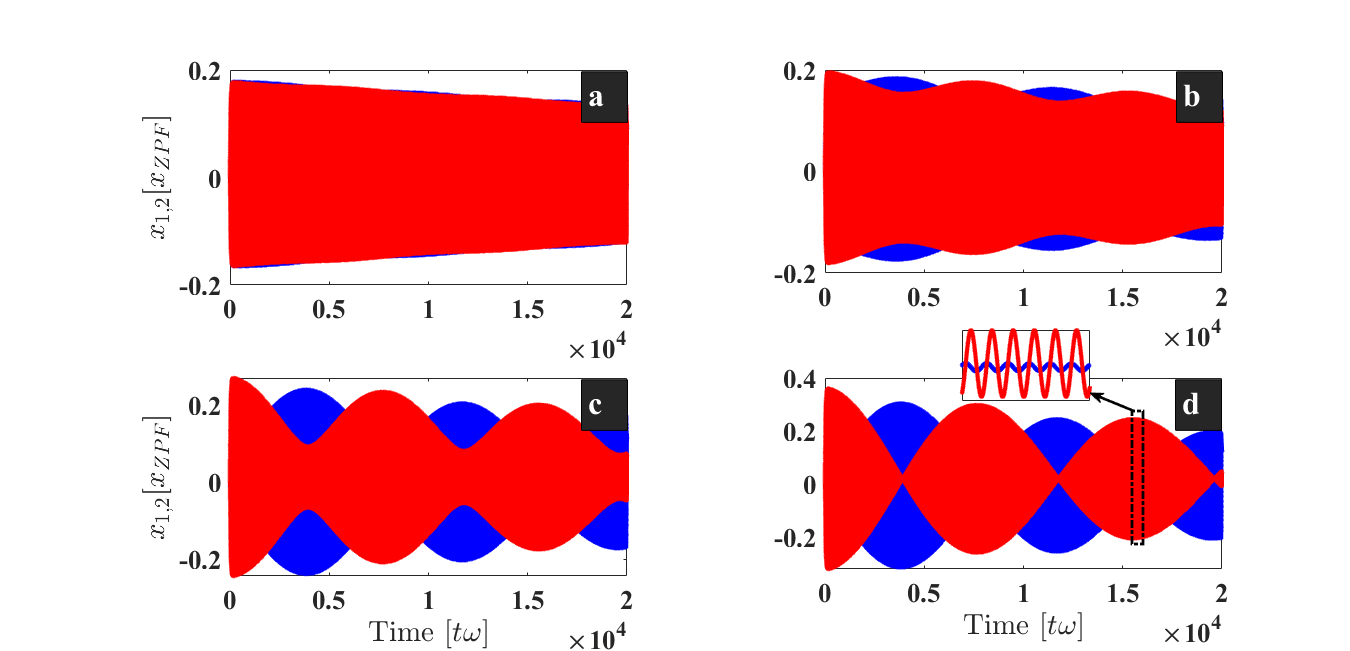}
\caption{Dynamical evolution of the mechanical resonators in the unbroken $\mathcal{PT}-$symmetric regime. (a)-(d) are the mechanical displacements for different values of the dissipative coupling rate, i.e. $\eta=0; 0.1; 0.5; g_m$. The driving force is fixed at $\alpha^{in}=20[\omega_m^{1/2}]$, the mechanical dissipation at $\gamma_m=1.076\times 10^{-5}\omega_m$ and the phonon hopping rate at $J_m=4\times10^{-4}\omega_m$. The other parameters are as in Fig.~\ref{fig:Fig2}.}
\label{fig:Fig4}
\end{figure}

\begin{figure}[tbh]
  \centering
  \includegraphics[width=.8\linewidth]{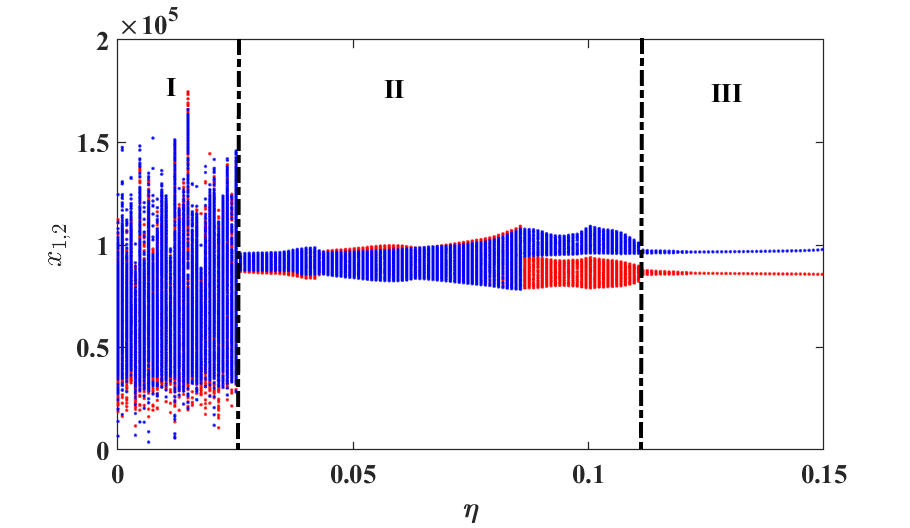}
  \caption{Bifurcation diagram showing the vibrational states of the mechanical resonators as a function of the dissipative coupling strength $\eta$. The driving strength is fixed at $\alpha^{in}=10^4[\omega_m^{1/2}]$, and the other parameters are as in Fig.~\ref{fig:Fig2}.}
  \label{fig:Fig5}
\end{figure}

To illustrate the effect of the dissipative coupling in our proposal, we first provide a general view of the dynamical behavior through a bifurcation diagram as shown in Fig.~\ref{fig:Fig5}, where the dynamics involved will be explained later.  This bifurcation diagram shows how our system exhibits chaotic dynamics for zero and weak values of the dissipative coupling ($\eta \in [0-0.025]\omega_m$). As the dissipative coupling increases, the system enters a quasi-periodic-like behavior ($\eta \in [0.026-0.111]\omega_m$), where regular dynamics are gradually recovered.  For sufficiently strong $\eta$, the chaotic dynamics is completely switched off and the regular dynamics is restored ($\eta>0.111\omega_m$). These three dynamical states are denoted as zones I, II and III in Fig.~\ref{fig:Fig5}, and are delineated by the dashed vertical lines to guide the eye.

To illustrate the effect of dissipative coupling in our proposal, we first provide a general view of the dynamical behavior through a bifurcation diagram as shown in Fig.~\ref{fig:Fig5}, where the dynamics involved will be explained later. This bifurcation diagram shows how our system exhibits chaotic dynamics for weak values of the dissipative coupling ($\eta \in [0-0.025]\omega_m$). As the dissipative coupling increases, the system enters a quasi-periodic-like behavior ($\eta \in [0.026-0.111]\omega_m$), where regular dynamics are gradually recovered.  For sufficiently large $\eta$, the chaotic dynamics is completely switched off and the regular dynamics is restored ($\eta>0.111\omega_m$). These three dynamical states are denoted as zones I, II and III in Fig.~\ref{fig:Fig5}, and are delimited by the dotted vertical lines to guide the eye.

\begin{figure}[tbhp]
  \centering
  \includegraphics[width=\linewidth]{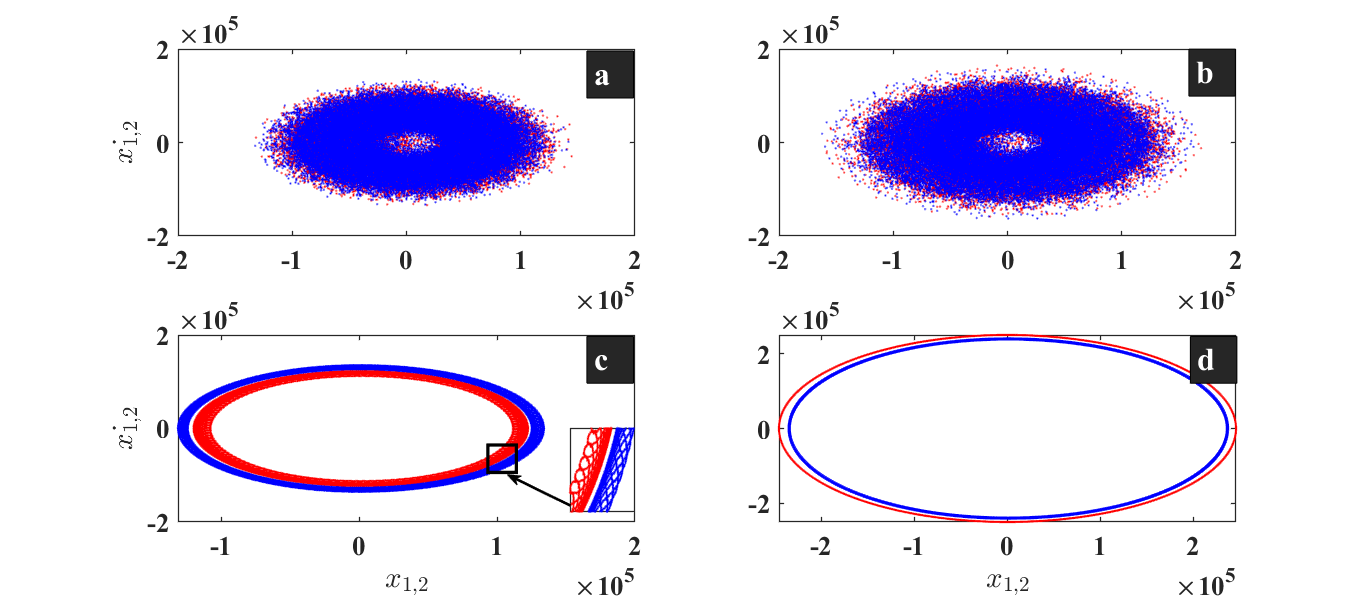}
  \caption{Poincaré section highlighting certain vibrational states based on the bifurcation diagram shown in Fig.~\ref{fig:Fig5}. The strengths of the dissipative coupling are $\tfrac{\eta}{g_m}= 0; 2\times10^{-2}; 9.7\times10^{-2}$; and $\tfrac{\eta}{g_m}=15.3\times10^{-2}$, respectively. The driving force is fixed at $\alpha^{in}=10^4[\omega_m^{1/2}]$, and the other parameters are as in Fig.~\ref{fig:Fig2}.}
  \label{fig:Fig6}
\end{figure}

To further highlight these dynamics, we have plotted the Poincaré section for certain values of $\eta$ in Fig.~\ref{fig:Fig6}. The dissipative coupling values used are $\eta= 0$, $\tfrac{\eta}{\omega_m}= 2\times10^{-2}$, $\tfrac{\eta}{\omega_m}= 9.7\times10^{-2}$, and $\tfrac{\eta}{\omega_m}= 15.3$, respectively. One can see that these Poincaré sections agree well with the dynamical state carried out in the bifurcation diagram. In fact, Fig.~\ref{fig:Fig6}(a,b) show a chaotic behavior in our system revealed by an abundance of random points, Fig.~\ref{fig:Fig6}c shows a quasi-periodic state, while Fig.~\ref{fig:Fig6}d reveals a regular dynamics in our system. 

\begin{figure}[tbhp]
  \centering
  \includegraphics[width=\linewidth]{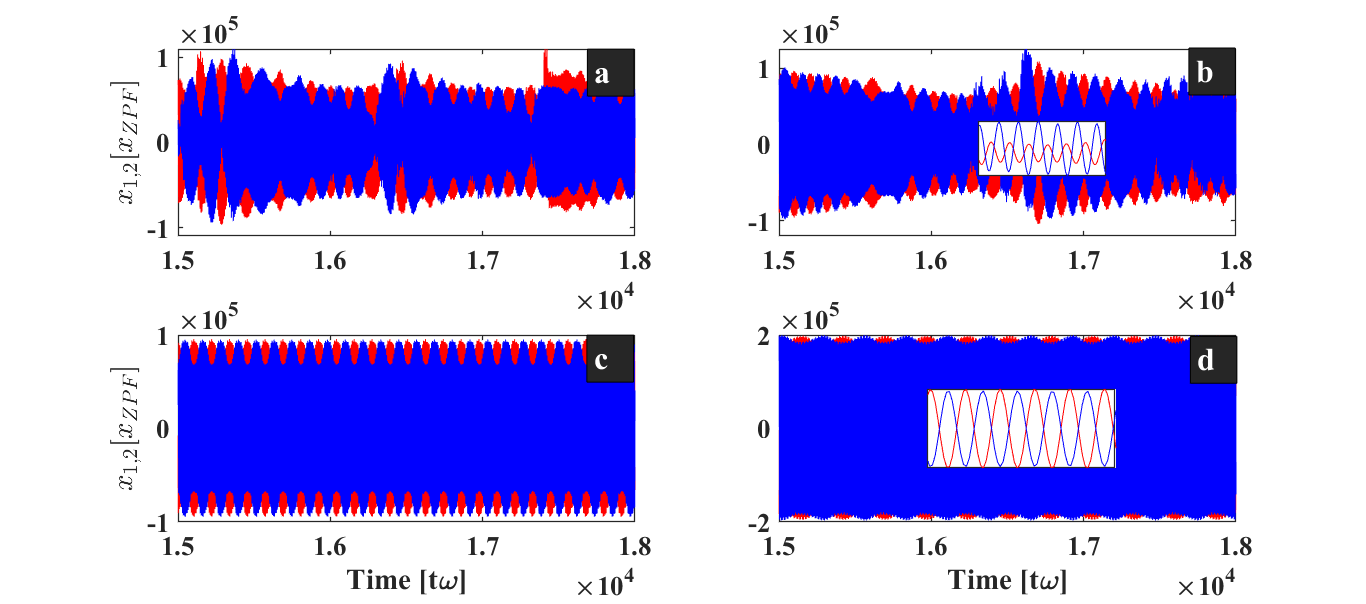}
  \caption{Dynamical evolution of the mechanical resonators in the broken $\mathcal{PT}-$symmetric regime. (a) and (b) show complex dynamics, and these complex dynamics are swapped out in (c) and (d), where regular dynamics are gradually restored. From (a) to (d), the dissipative coupling strengths correspond to those in Fig.~\ref{fig:Fig6}(a) to Fig.~\ref{fig:Fig6}(d), respectively. The other parameters are the same as in Fig.~\ref{fig:Fig2}.}
  \label{fig:Fig7}
\end{figure}

For more details about these dynamical states, we have plotted their stationary time evolution in Fig.~\ref{fig:Fig7}. It can be seen that Fig.~\ref{fig:Fig7}(a-b) exhibit complex dynamics as predicted both in the bifurcation diagram and in the Poincaré sections shown in Fig.~\ref{fig:Fig5} and Fig.~\ref{fig:Fig6}, respectively. Such complex dynamics, resulting from a distortion of the oscillation beats induced by a strong coupling in our system, are referred to as chaotic beats \cite{Ahamed_2011}. Furthermore, the regular dynamical states carried out in Fig.~\ref{fig:Fig7}(c-d) are also in agreement with the above predictions. These analyses show how dissipative coupling can be used to switch from a chaotic to a regular state. This dynamical control shows how $\eta$ can be used to suppress unwanted behavior in dynamical systems. Moreover, the oscillation beats shown in Fig.~\ref{fig:Fig7}c suggest how the dissipative coupling tends to restore $\mathcal{PT}-$symmetry in the system, which is similar to the investigation carried out in Fig.~\ref{fig:Fig4}.    

\section{Conclusion} \label{sec:concl} 

We have investigated the effect of the dissipative coupling on a mechanically coupled optomechanical system. In our proposal, one cavity is driven with blue-detuned field to induce gain, while the other is driven with red-detuned field to induce losses. These gain and loss rates are shown to depend on the dissipative coupling parameter. We have shown that EP occurs at low driving strengths when dissipative coupling is turned on, compared to a purely dispersive system. In the linear regime, where the $\mathcal{PT}-$ symmetry is unbroken, the dissipative coupling induces a strong coupling between the mechanical resonators, leading to an enhancement of the energy exchange between them. Above the parametric instability threshold, which occurs at the EP, the $\mathcal{PT}-$symmetry is broken and the system enters into a nonlinear regime. In this broken $\mathcal{PT}-$symmetry regime, the purely dispersive system exhibits complex dynamics represented by chaotic beats. These complex dynamics are controlled by the dissipative coupling, which switches the chaotic beats into regular beat oscillations. These results show how dissipative $\mathcal{PT}-$symmetry systems can be used to harness new features of EP with less driving power compared to their dispersive counterparts. Furthermore, this work extends the EP opportunities to dissipative optomechanical systems, and paves the way to study $\mathcal{PT}-$symmetry features at low-threshold power.

\section*{Acknowledgments}
This work was carried out under the Iso-Lomso Fellowship at the Stellenbosch Institute for Advanced Study (STIAS), Wallenberg Research Centre at Stellenbosch University, Stellenbosch 7600, South Africa. 

P. Djorwe and S.G. Nana Engo thank the Ministry of Higher Education of Cameroon (MINESUP) for the financial support within the framework of the "Research Modernization" grants.

\bibliography{Sensor}

\begin{thebibliography}{36}
\expandafter\ifx\csname natexlab\endcsname\relax\def\natexlab#1{#1}\fi
\expandafter\ifx\csname bibnamefont\endcsname\relax
  \def\bibnamefont#1{#1}\fi
\expandafter\ifx\csname bibfnamefont\endcsname\relax
  \def\bibfnamefont#1{#1}\fi
\expandafter\ifx\csname citenamefont\endcsname\relax
  \def\citenamefont#1{#1}\fi
\expandafter\ifx\csname url\endcsname\relax
  \def\url#1{\texttt{#1}}\fi
\expandafter\ifx\csname urlprefix\endcsname\relax\def\urlprefix{URL }\fi
\providecommand{\bibinfo}[2]{#2}
\providecommand{\eprint}[2][]{\url{#2}}

\bibitem[{\citenamefont{Aspelmeyer et~al.}(2014)\citenamefont{Aspelmeyer,
  Kippenberg, and Marquardt}}]{Aspelmeyer_2014}
\bibinfo{author}{\bibfnamefont{M.}~\bibnamefont{Aspelmeyer}},
  \bibinfo{author}{\bibfnamefont{T.~J.} \bibnamefont{Kippenberg}},
  \bibnamefont{and}
  \bibinfo{author}{\bibfnamefont{F.}~\bibnamefont{Marquardt}},
  \bibinfo{journal}{Reviews of Modern Physics} \textbf{\bibinfo{volume}{86}},
  \bibinfo{pages}{1391} (\bibinfo{year}{2014}).

\bibitem[{\citenamefont{Li et~al.}(2009)\citenamefont{Li, Pernice, and
  Tang}}]{Li_2009}
\bibinfo{author}{\bibfnamefont{M.}~\bibnamefont{Li}},
  \bibinfo{author}{\bibfnamefont{W.~H.~P.} \bibnamefont{Pernice}},
  \bibnamefont{and} \bibinfo{author}{\bibfnamefont{H.~X.} \bibnamefont{Tang}},
  \bibinfo{journal}{Physical Review Letters} \textbf{\bibinfo{volume}{103}},
  \bibinfo{pages}{223901} (\bibinfo{year}{2009}).

\bibitem[{\citenamefont{Huang and Agarwal}(2010)}]{Huang_2010}
\bibinfo{author}{\bibfnamefont{S.}~\bibnamefont{Huang}} \bibnamefont{and}
  \bibinfo{author}{\bibfnamefont{G.~S.} \bibnamefont{Agarwal}},
  \bibinfo{journal}{Physical Review A} \textbf{\bibinfo{volume}{81}},
  \bibinfo{pages}{053810} (\bibinfo{year}{2010}).

\bibitem[{\citenamefont{Fu et~al.}(2012)\citenamefont{Fu, Gu, Yan, Yang, Cui,
  and Wu}}]{Fu_2012}
\bibinfo{author}{\bibfnamefont{C.-B.} \bibnamefont{Fu}},
  \bibinfo{author}{\bibfnamefont{K.-H.} \bibnamefont{Gu}},
  \bibinfo{author}{\bibfnamefont{X.-B.} \bibnamefont{Yan}},
  \bibinfo{author}{\bibfnamefont{X.}~\bibnamefont{Yang}},
  \bibinfo{author}{\bibfnamefont{C.-L.} \bibnamefont{Cui}}, \bibnamefont{and}
  \bibinfo{author}{\bibfnamefont{J.-H.} \bibnamefont{Wu}},
  \bibinfo{journal}{Physics Letters A} \textbf{\bibinfo{volume}{377}},
  \bibinfo{pages}{133} (\bibinfo{year}{2012}).

\bibitem[{\citenamefont{Weiss et~al.}(2013)\citenamefont{Weiss, Bruder, and
  Nunnenkamp}}]{Weiss_2013}
\bibinfo{author}{\bibfnamefont{T.}~\bibnamefont{Weiss}},
  \bibinfo{author}{\bibfnamefont{C.}~\bibnamefont{Bruder}}, \bibnamefont{and}
  \bibinfo{author}{\bibfnamefont{A.}~\bibnamefont{Nunnenkamp}},
  \bibinfo{journal}{New Journal of Physics} \textbf{\bibinfo{volume}{15}},
  \bibinfo{pages}{045017} (\bibinfo{year}{2013}).

\bibitem[{\citenamefont{ju~Gu et~al.}(2013)\citenamefont{ju~Gu, xiang Li, and
  ping Yang}}]{Gu_2013}
\bibinfo{author}{\bibfnamefont{W.}~\bibnamefont{ju~Gu}},
  \bibinfo{author}{\bibfnamefont{G.}~\bibnamefont{xiang Li}}, \bibnamefont{and}
  \bibinfo{author}{\bibfnamefont{Y.}~\bibnamefont{ping Yang}},
  \bibinfo{journal}{Physical Review A} \textbf{\bibinfo{volume}{88}},
  \bibinfo{pages}{013835} (\bibinfo{year}{2013}).

\bibitem[{\citenamefont{Qu and Agarwal}(2015)}]{Qu_2015}
\bibinfo{author}{\bibfnamefont{K.}~\bibnamefont{Qu}} \bibnamefont{and}
  \bibinfo{author}{\bibfnamefont{G.~S.} \bibnamefont{Agarwal}},
  \bibinfo{journal}{Physical Review A} \textbf{\bibinfo{volume}{91}},
  \bibinfo{pages}{063815} (\bibinfo{year}{2015}).

\bibitem[{\citenamefont{Wu et~al.}(2014)\citenamefont{Wu, Hryciw, Healey, Lake,
  Jayakumar, Freeman, Davis, and Barclay}}]{Wu_2014}
\bibinfo{author}{\bibfnamefont{M.}~\bibnamefont{Wu}},
  \bibinfo{author}{\bibfnamefont{A.~C.} \bibnamefont{Hryciw}},
  \bibinfo{author}{\bibfnamefont{C.}~\bibnamefont{Healey}},
  \bibinfo{author}{\bibfnamefont{D.~P.} \bibnamefont{Lake}},
  \bibinfo{author}{\bibfnamefont{H.}~\bibnamefont{Jayakumar}},
  \bibinfo{author}{\bibfnamefont{M.~R.} \bibnamefont{Freeman}},
  \bibinfo{author}{\bibfnamefont{J.~P.} \bibnamefont{Davis}}, \bibnamefont{and}
  \bibinfo{author}{\bibfnamefont{P.~E.} \bibnamefont{Barclay}},
  \bibinfo{journal}{Physical Review X} \textbf{\bibinfo{volume}{4}},
  \bibinfo{pages}{021052} (\bibinfo{year}{2014}).

\bibitem[{\citenamefont{Tagantsev et~al.}(2018)\citenamefont{Tagantsev,
  Sokolov, and Polzik}}]{Tagantsev_2018}
\bibinfo{author}{\bibfnamefont{A.~K.} \bibnamefont{Tagantsev}},
  \bibinfo{author}{\bibfnamefont{I.~V.} \bibnamefont{Sokolov}},
  \bibnamefont{and} \bibinfo{author}{\bibfnamefont{E.~S.}
  \bibnamefont{Polzik}}, \bibinfo{journal}{Physical Review A}
  \textbf{\bibinfo{volume}{97}}, \bibinfo{pages}{063820}
  (\bibinfo{year}{2018}).

\bibitem[{\citenamefont{Clark et~al.}(2017)\citenamefont{Clark, Lecocq,
  Simmonds, Aumentado, and Teufel}}]{Clark_2017}
\bibinfo{author}{\bibfnamefont{J.~B.} \bibnamefont{Clark}},
  \bibinfo{author}{\bibfnamefont{F.}~\bibnamefont{Lecocq}},
  \bibinfo{author}{\bibfnamefont{R.~W.} \bibnamefont{Simmonds}},
  \bibinfo{author}{\bibfnamefont{J.}~\bibnamefont{Aumentado}},
  \bibnamefont{and} \bibinfo{author}{\bibfnamefont{J.~D.}
  \bibnamefont{Teufel}}, \bibinfo{journal}{Nature}
  \textbf{\bibinfo{volume}{541}}, \bibinfo{pages}{191} (\bibinfo{year}{2017}).

\bibitem[{\citenamefont{Djorw{\'{e}} et~al.}(2012)\citenamefont{Djorw{\'{e}},
  Mb{\'{e}}, Engo, and Woafo}}]{Djorw__2012}
\bibinfo{author}{\bibfnamefont{P.}~\bibnamefont{Djorw{\'{e}}}},
  \bibinfo{author}{\bibfnamefont{J.~H.~T.} \bibnamefont{Mb{\'{e}}}},
  \bibinfo{author}{\bibfnamefont{S.~G.~N.} \bibnamefont{Engo}},
  \bibnamefont{and} \bibinfo{author}{\bibfnamefont{P.}~\bibnamefont{Woafo}},
  \bibinfo{journal}{Physical Review A} \textbf{\bibinfo{volume}{86}},
  \bibinfo{pages}{043816} (\bibinfo{year}{2012}).

\bibitem[{\citenamefont{Kotler et~al.}(2021)\citenamefont{Kotler, Peterson,
  Shojaee, Lecocq, Cicak, Kwiatkowski, Geller, Glancy, Knill, Simmonds
  et~al.}}]{Kotler_2021}
\bibinfo{author}{\bibfnamefont{S.}~\bibnamefont{Kotler}},
  \bibinfo{author}{\bibfnamefont{G.~A.} \bibnamefont{Peterson}},
  \bibinfo{author}{\bibfnamefont{E.}~\bibnamefont{Shojaee}},
  \bibinfo{author}{\bibfnamefont{F.}~\bibnamefont{Lecocq}},
  \bibinfo{author}{\bibfnamefont{K.}~\bibnamefont{Cicak}},
  \bibinfo{author}{\bibfnamefont{A.}~\bibnamefont{Kwiatkowski}},
  \bibinfo{author}{\bibfnamefont{S.}~\bibnamefont{Geller}},
  \bibinfo{author}{\bibfnamefont{S.}~\bibnamefont{Glancy}},
  \bibinfo{author}{\bibfnamefont{E.}~\bibnamefont{Knill}},
  \bibinfo{author}{\bibfnamefont{R.~W.} \bibnamefont{Simmonds}},
  \bibnamefont{et~al.}, \bibinfo{journal}{Science}
  \textbf{\bibinfo{volume}{372}}, \bibinfo{pages}{622} (\bibinfo{year}{2021}).

\bibitem[{\citenamefont{Tchodimou et~al.}(2017)\citenamefont{Tchodimou, Djorwe,
  and Engo}}]{Tchodimou_2017}
\bibinfo{author}{\bibfnamefont{C.}~\bibnamefont{Tchodimou}},
  \bibinfo{author}{\bibfnamefont{P.}~\bibnamefont{Djorwe}}, \bibnamefont{and}
  \bibinfo{author}{\bibfnamefont{S.~G.~N.} \bibnamefont{Engo}},
  \bibinfo{journal}{Physical Review A} \textbf{\bibinfo{volume}{96}},
  \bibinfo{pages}{033856} (\bibinfo{year}{2017}).

\bibitem[{\citenamefont{Djorw{\'{e}} et~al.}(2014)\citenamefont{Djorw{\'{e}},
  Engo, and Woafo}}]{Djorw__2014}
\bibinfo{author}{\bibfnamefont{P.}~\bibnamefont{Djorw{\'{e}}}},
  \bibinfo{author}{\bibfnamefont{S.~G.~N.} \bibnamefont{Engo}},
  \bibnamefont{and} \bibinfo{author}{\bibfnamefont{P.}~\bibnamefont{Woafo}},
  \bibinfo{journal}{Physical Review A} \textbf{\bibinfo{volume}{90}},
  \bibinfo{pages}{024303} (\bibinfo{year}{2014}).

\bibitem[{\citenamefont{Wollman et~al.}(2015)\citenamefont{Wollman, Lei,
  Weinstein, Suh, Kronwald, Marquardt, Clerk, and Schwab}}]{Wollman_2015}
\bibinfo{author}{\bibfnamefont{E.~E.} \bibnamefont{Wollman}},
  \bibinfo{author}{\bibfnamefont{C.~U.} \bibnamefont{Lei}},
  \bibinfo{author}{\bibfnamefont{A.~J.} \bibnamefont{Weinstein}},
  \bibinfo{author}{\bibfnamefont{J.}~\bibnamefont{Suh}},
  \bibinfo{author}{\bibfnamefont{A.}~\bibnamefont{Kronwald}},
  \bibinfo{author}{\bibfnamefont{F.}~\bibnamefont{Marquardt}},
  \bibinfo{author}{\bibfnamefont{A.~A.} \bibnamefont{Clerk}}, \bibnamefont{and}
  \bibinfo{author}{\bibfnamefont{K.~C.} \bibnamefont{Schwab}},
  \bibinfo{journal}{Science} \textbf{\bibinfo{volume}{349}},
  \bibinfo{pages}{952} (\bibinfo{year}{2015}).

\bibitem[{\citenamefont{Djorw{\'{e}}
  et~al.}(2013{\natexlab{a}})\citenamefont{Djorw{\'{e}}, Engo, Mb{\'{e}}, and
  Woafo}}]{Djorwe2013}
\bibinfo{author}{\bibfnamefont{P.}~\bibnamefont{Djorw{\'{e}}}},
  \bibinfo{author}{\bibfnamefont{S.~N.} \bibnamefont{Engo}},
  \bibinfo{author}{\bibfnamefont{J.~T.} \bibnamefont{Mb{\'{e}}}},
  \bibnamefont{and} \bibinfo{author}{\bibfnamefont{P.}~\bibnamefont{Woafo}},
  \bibinfo{journal}{Physica B: Condensed Matter}
  \textbf{\bibinfo{volume}{422}}, \bibinfo{pages}{72}
  (\bibinfo{year}{2013}{\natexlab{a}}).

\bibitem[{\citenamefont{Alphonse et~al.}(2022)\citenamefont{Alphonse, Djorwe,
  Abbagari, Doka, and Engo}}]{Alphonse_2022}
\bibinfo{author}{\bibfnamefont{H.}~\bibnamefont{Alphonse}},
  \bibinfo{author}{\bibfnamefont{P.}~\bibnamefont{Djorwe}},
  \bibinfo{author}{\bibfnamefont{S.}~\bibnamefont{Abbagari}},
  \bibinfo{author}{\bibfnamefont{S.~Y.} \bibnamefont{Doka}}, \bibnamefont{and}
  \bibinfo{author}{\bibfnamefont{S.~N.} \bibnamefont{Engo}},
  \bibinfo{journal}{Chaos, Solitons and Fractals}
  \textbf{\bibinfo{volume}{154}}, \bibinfo{pages}{111593}
  (\bibinfo{year}{2022}).

\bibitem[{\citenamefont{Djorwe et~al.}(2019{\natexlab{a}})\citenamefont{Djorwe,
  Pennec, and Djafari-Rouhani}}]{Djorwe2019}
\bibinfo{author}{\bibfnamefont{P.}~\bibnamefont{Djorwe}},
  \bibinfo{author}{\bibfnamefont{Y.}~\bibnamefont{Pennec}}, \bibnamefont{and}
  \bibinfo{author}{\bibfnamefont{B.}~\bibnamefont{Djafari-Rouhani}},
  \bibinfo{journal}{Scientific Reports} \textbf{\bibinfo{volume}{9}},
  \bibinfo{pages}{1684} (\bibinfo{year}{2019}{\natexlab{a}}).

\bibitem[{\citenamefont{Djorw{\'{e}} et~al.}(2020)\citenamefont{Djorw{\'{e}},
  Pennec, and Djafari-Rouhani}}]{Djorwe_2020}
\bibinfo{author}{\bibfnamefont{P.}~\bibnamefont{Djorw{\'{e}}}},
  \bibinfo{author}{\bibfnamefont{Y.}~\bibnamefont{Pennec}}, \bibnamefont{and}
  \bibinfo{author}{\bibfnamefont{B.}~\bibnamefont{Djafari-Rouhani}},
  \bibinfo{journal}{Physical Review B} \textbf{\bibinfo{volume}{102}},
  \bibinfo{pages}{155410} (\bibinfo{year}{2020}).

\bibitem[{\citenamefont{Monifi et~al.}(2016)\citenamefont{Monifi, Zhang,
  Özdemir, Peng, xi~Liu, Bo, Nori, and Yang}}]{Monifi_2016}
\bibinfo{author}{\bibfnamefont{F.}~\bibnamefont{Monifi}},
  \bibinfo{author}{\bibfnamefont{J.}~\bibnamefont{Zhang}},
  \bibinfo{author}{\bibfnamefont{{\c{S}}.~K.} \bibnamefont{Özdemir}},
  \bibinfo{author}{\bibfnamefont{B.}~\bibnamefont{Peng}},
  \bibinfo{author}{\bibfnamefont{Y.}~\bibnamefont{xi~Liu}},
  \bibinfo{author}{\bibfnamefont{F.}~\bibnamefont{Bo}},
  \bibinfo{author}{\bibfnamefont{F.}~\bibnamefont{Nori}}, \bibnamefont{and}
  \bibinfo{author}{\bibfnamefont{L.}~\bibnamefont{Yang}},
  \bibinfo{journal}{Nature Photonics} \textbf{\bibinfo{volume}{10}},
  \bibinfo{pages}{399} (\bibinfo{year}{2016}).

\bibitem[{\citenamefont{Djorwe et~al.}(2018)\citenamefont{Djorwe, Pennec, and
  Djafari-Rouhani}}]{Djorwe_2018}
\bibinfo{author}{\bibfnamefont{P.}~\bibnamefont{Djorwe}},
  \bibinfo{author}{\bibfnamefont{Y.}~\bibnamefont{Pennec}}, \bibnamefont{and}
  \bibinfo{author}{\bibfnamefont{B.}~\bibnamefont{Djafari-Rouhani}},
  \bibinfo{journal}{Physical Review E} \textbf{\bibinfo{volume}{98}},
  \bibinfo{pages}{032201} (\bibinfo{year}{2018}).

\bibitem[{\citenamefont{Kingni et~al.}(2020)\citenamefont{Kingni, Tchodimou,
  Foulla, Djorwe, and Engo}}]{Kingni_2020}
\bibinfo{author}{\bibfnamefont{S.~T.} \bibnamefont{Kingni}},
  \bibinfo{author}{\bibfnamefont{C.}~\bibnamefont{Tchodimou}},
  \bibinfo{author}{\bibfnamefont{D.~P.} \bibnamefont{Foulla}},
  \bibinfo{author}{\bibfnamefont{P.}~\bibnamefont{Djorwe}}, \bibnamefont{and}
  \bibinfo{author}{\bibfnamefont{S.~G.~N.} \bibnamefont{Engo}},
  \bibinfo{journal}{The European Physical Journal Special Topics}
  \textbf{\bibinfo{volume}{229}}, \bibinfo{pages}{1117} (\bibinfo{year}{2020}).

\bibitem[{\citenamefont{Miri and Alù}(2019)}]{Miri_2019}
\bibinfo{author}{\bibfnamefont{M.-A.} \bibnamefont{Miri}} \bibnamefont{and}
  \bibinfo{author}{\bibfnamefont{A.}~\bibnamefont{Alù}},
  \bibinfo{journal}{Science} \textbf{\bibinfo{volume}{363}},
  \bibinfo{pages}{eaar7709} (\bibinfo{year}{2019}), ISSN
  \bibinfo{issn}{0036-8075}.

\bibitem[{\citenamefont{Goldzak et~al.}(2018)\citenamefont{Goldzak, Mailybaev,
  and Moiseyev}}]{Goldzak_2018}
\bibinfo{author}{\bibfnamefont{T.}~\bibnamefont{Goldzak}},
  \bibinfo{author}{\bibfnamefont{A.~A.} \bibnamefont{Mailybaev}},
  \bibnamefont{and} \bibinfo{author}{\bibfnamefont{N.}~\bibnamefont{Moiseyev}},
  \bibinfo{journal}{Physical Review Letters} \textbf{\bibinfo{volume}{120}},
  \bibinfo{pages}{013901} (\bibinfo{year}{2018}).

\bibitem[{\citenamefont{Peng et~al.}(2014)\citenamefont{Peng, Özdemir, Rotter,
  Yilmaz, Liertzer, Monifi, Bender, Nori, and Yang}}]{Peng_2014}
\bibinfo{author}{\bibfnamefont{B.}~\bibnamefont{Peng}},
  \bibinfo{author}{\bibfnamefont{{\c{S}}.~K.} \bibnamefont{Özdemir}},
  \bibinfo{author}{\bibfnamefont{S.}~\bibnamefont{Rotter}},
  \bibinfo{author}{\bibfnamefont{H.}~\bibnamefont{Yilmaz}},
  \bibinfo{author}{\bibfnamefont{M.}~\bibnamefont{Liertzer}},
  \bibinfo{author}{\bibfnamefont{F.}~\bibnamefont{Monifi}},
  \bibinfo{author}{\bibfnamefont{C.~M.} \bibnamefont{Bender}},
  \bibinfo{author}{\bibfnamefont{F.}~\bibnamefont{Nori}}, \bibnamefont{and}
  \bibinfo{author}{\bibfnamefont{L.}~\bibnamefont{Yang}},
  \bibinfo{journal}{Science} \textbf{\bibinfo{volume}{346}},
  \bibinfo{pages}{328} (\bibinfo{year}{2014}).

\bibitem[{\citenamefont{Djorwe et~al.}(2019{\natexlab{b}})\citenamefont{Djorwe,
  Pennec, and Djafari-Rouhani}}]{Djorwe_2019}
\bibinfo{author}{\bibfnamefont{P.}~\bibnamefont{Djorwe}},
  \bibinfo{author}{\bibfnamefont{Y.}~\bibnamefont{Pennec}}, \bibnamefont{and}
  \bibinfo{author}{\bibfnamefont{B.}~\bibnamefont{Djafari-Rouhani}},
  \bibinfo{journal}{Physical Review Applied} \textbf{\bibinfo{volume}{12}},
  \bibinfo{pages}{024002} (\bibinfo{year}{2019}{\natexlab{b}}).

\bibitem[{\citenamefont{Chen et~al.}(2017)\citenamefont{Chen, Özdemir, Zhao,
  Wiersig, and Yang}}]{Chen_2017}
\bibinfo{author}{\bibfnamefont{W.}~\bibnamefont{Chen}},
  \bibinfo{author}{\bibfnamefont{{\c{S}}.~K.} \bibnamefont{Özdemir}},
  \bibinfo{author}{\bibfnamefont{G.}~\bibnamefont{Zhao}},
  \bibinfo{author}{\bibfnamefont{J.}~\bibnamefont{Wiersig}}, \bibnamefont{and}
  \bibinfo{author}{\bibfnamefont{L.}~\bibnamefont{Yang}},
  \bibinfo{journal}{Nature} \textbf{\bibinfo{volume}{548}},
  \bibinfo{pages}{192} (\bibinfo{year}{2017}).

\bibitem[{\citenamefont{Hodaei et~al.}(2017)\citenamefont{Hodaei, Hassan,
  Wittek, Garcia-Gracia, El-Ganainy, Christodoulides, and
  Khajavikhan}}]{Hodaei_2017}
\bibinfo{author}{\bibfnamefont{H.}~\bibnamefont{Hodaei}},
  \bibinfo{author}{\bibfnamefont{A.~U.} \bibnamefont{Hassan}},
  \bibinfo{author}{\bibfnamefont{S.}~\bibnamefont{Wittek}},
  \bibinfo{author}{\bibfnamefont{H.}~\bibnamefont{Garcia-Gracia}},
  \bibinfo{author}{\bibfnamefont{R.}~\bibnamefont{El-Ganainy}},
  \bibinfo{author}{\bibfnamefont{D.~N.} \bibnamefont{Christodoulides}},
  \bibnamefont{and}
  \bibinfo{author}{\bibfnamefont{M.}~\bibnamefont{Khajavikhan}},
  \bibinfo{journal}{Nature} \textbf{\bibinfo{volume}{548}},
  \bibinfo{pages}{187} (\bibinfo{year}{2017}).

\bibitem[{\citenamefont{Xu et~al.}(2016)\citenamefont{Xu, Mason, Jiang, and
  Harris}}]{Xu_2016}
\bibinfo{author}{\bibfnamefont{H.}~\bibnamefont{Xu}},
  \bibinfo{author}{\bibfnamefont{D.}~\bibnamefont{Mason}},
  \bibinfo{author}{\bibfnamefont{L.}~\bibnamefont{Jiang}}, \bibnamefont{and}
  \bibinfo{author}{\bibfnamefont{J.~G.~E.} \bibnamefont{Harris}},
  \bibinfo{journal}{Nature} \textbf{\bibinfo{volume}{537}}, \bibinfo{pages}{80}
  (\bibinfo{year}{2016}).

\bibitem[{\citenamefont{Ren et~al.}(2022)\citenamefont{Ren, Shah, Pfeifer,
  Brendel, Peano, Marquardt, and Painter}}]{Ren_2022}
\bibinfo{author}{\bibfnamefont{H.}~\bibnamefont{Ren}},
  \bibinfo{author}{\bibfnamefont{T.}~\bibnamefont{Shah}},
  \bibinfo{author}{\bibfnamefont{H.}~\bibnamefont{Pfeifer}},
  \bibinfo{author}{\bibfnamefont{C.}~\bibnamefont{Brendel}},
  \bibinfo{author}{\bibfnamefont{V.}~\bibnamefont{Peano}},
  \bibinfo{author}{\bibfnamefont{F.}~\bibnamefont{Marquardt}},
  \bibnamefont{and} \bibinfo{author}{\bibfnamefont{O.}~\bibnamefont{Painter}},
  \bibinfo{journal}{Nature Communications} \textbf{\bibinfo{volume}{13}},
  \bibinfo{eid}{3476} (\bibinfo{year}{2022}), ISSN \bibinfo{issn}{2041-1723}.

\bibitem[{\citenamefont{Peano et~al.}(2015)\citenamefont{Peano, Brendel,
  Schmidt, and Marquardt}}]{Peano_2015}
\bibinfo{author}{\bibfnamefont{V.}~\bibnamefont{Peano}},
  \bibinfo{author}{\bibfnamefont{C.}~\bibnamefont{Brendel}},
  \bibinfo{author}{\bibfnamefont{M.}~\bibnamefont{Schmidt}}, \bibnamefont{and}
  \bibinfo{author}{\bibfnamefont{F.}~\bibnamefont{Marquardt}},
  \bibinfo{journal}{Physical Review X} \textbf{\bibinfo{volume}{5}},
  \bibinfo{pages}{031011} (\bibinfo{year}{2015}).

\bibitem[{\citenamefont{Foulla et~al.}(2017)\citenamefont{Foulla, Djorw{\'{e}},
  Kingni, and Engo}}]{Foulla_2017}
\bibinfo{author}{\bibfnamefont{D.~P.} \bibnamefont{Foulla}},
  \bibinfo{author}{\bibfnamefont{P.}~\bibnamefont{Djorw{\'{e}}}},
  \bibinfo{author}{\bibfnamefont{S.~T.} \bibnamefont{Kingni}},
  \bibnamefont{and} \bibinfo{author}{\bibfnamefont{S.~G.~N.}
  \bibnamefont{Engo}}, \bibinfo{journal}{Physical Review A}
  \textbf{\bibinfo{volume}{95}}, \bibinfo{pages}{013831}
  (\bibinfo{year}{2017}).

\bibitem[{\citenamefont{Wang et~al.}(2014)\citenamefont{Wang, Huang, Lai, and
  Grebogi}}]{Wang_2014}
\bibinfo{author}{\bibfnamefont{G.}~\bibnamefont{Wang}},
  \bibinfo{author}{\bibfnamefont{L.}~\bibnamefont{Huang}},
  \bibinfo{author}{\bibfnamefont{Y.-C.} \bibnamefont{Lai}}, \bibnamefont{and}
  \bibinfo{author}{\bibfnamefont{C.}~\bibnamefont{Grebogi}},
  \bibinfo{journal}{Physical Review Letters} \textbf{\bibinfo{volume}{112}},
  \bibinfo{pages}{110406} (\bibinfo{year}{2014}).

\bibitem[{\citenamefont{Jiang et~al.}(2021)\citenamefont{Jiang, Liu, and
  Sillanpää}}]{Jiang_2021}
\bibinfo{author}{\bibfnamefont{C.}~\bibnamefont{Jiang}},
  \bibinfo{author}{\bibfnamefont{Y.-L.} \bibnamefont{Liu}}, \bibnamefont{and}
  \bibinfo{author}{\bibfnamefont{M.~A.} \bibnamefont{Sillanpää}},
  \bibinfo{journal}{Physical Review A} \textbf{\bibinfo{volume}{104}},
  \bibinfo{pages}{013502} (\bibinfo{year}{2021}).

\bibitem[{\citenamefont{Djorw{\'{e}}
  et~al.}(2013{\natexlab{b}})\citenamefont{Djorw{\'{e}}, Mb{\'{e}}, Engo, and
  Woafo}}]{Djorwe.2013}
\bibinfo{author}{\bibfnamefont{P.}~\bibnamefont{Djorw{\'{e}}}},
  \bibinfo{author}{\bibfnamefont{J.~H.~T.} \bibnamefont{Mb{\'{e}}}},
  \bibinfo{author}{\bibfnamefont{S.~G.~N.} \bibnamefont{Engo}},
  \bibnamefont{and} \bibinfo{author}{\bibfnamefont{P.}~\bibnamefont{Woafo}},
  \bibinfo{journal}{The European Physical Journal D}
  \textbf{\bibinfo{volume}{67}}, \bibinfo{pages}{45}
  (\bibinfo{year}{2013}{\natexlab{b}}).

\bibitem[{\citenamefont{Ahamed et~al.}(2011)\citenamefont{Ahamed, Srinivansan,
  Murali, and Lakshmanan}}]{Ahamed_2011}
\bibinfo{author}{\bibfnamefont{A.~I.} \bibnamefont{Ahamed}},
  \bibinfo{author}{\bibfnamefont{K.}~\bibnamefont{Srinivansan}},
  \bibinfo{author}{\bibfnamefont{K.}~\bibnamefont{Murali}}, \bibnamefont{and}
  \bibinfo{author}{\bibfnamefont{M.}~\bibnamefont{Lakshmanan}},
  \bibinfo{journal}{International Journal of Bifurcation and Chaos}
  \textbf{\bibinfo{volume}{21}}, \bibinfo{pages}{737} (\bibinfo{year}{2011}).

\end{thebibliography}

\end{document}